\documentclass[%
 reprint,
superscriptaddress,
 amsmath,amssymb,
 aps,
 pre,
floatfix,
]{revtex4-2}
\usepackage{multirow}
\usepackage{graphicx}% Include figure files
\usepackage{dcolumn}% Align table columns on decimal point
\usepackage{bm}% bold math
\usepackage{xcolor}
\usepackage{soul}
\usepackage[normalem]{ulem}

\soulregister{\cite}{1}
\soulregister{\ref}{1}
\begin{document}

\title{Spectral Diagnostics of Higher-Order Structural Reorganization in Finite Statistical Systems}% Force line breaks with \\

\author{Wei Liu}
 \email{weiliu@xust.edu.cn}
 \affiliation{%
 College of Sciences, Xi'an University of Science and Technology, Xi'an 710054, China
}%
\author{Songzhi Lv}
\affiliation{%
 College of Sciences, Xi'an University of Science and Technology, Xi'an 710054, China
}%
\author{Xin Zhang}%
\affiliation{%
 College of Sciences, Xi'an University of Science and Technology, Xi'an 710054, China
}%

\author{Fangfang Wang}
\affiliation{
 School of Systems Science, Beijing Normal University, Beijing 100875, China
}
\author{Kai Qi}
\affiliation{2020 X-Lab, Shanghai Institute of Microsystems and Information Technology, Chinese Academy of Sciences, Shanghai 200050, China} 
\author{Zengru Di}
\email{zdi@bnu.edu.cn}
\affiliation{
 School of Systems Science, Beijing Normal University, Beijing 100875, China
}%

\begin{abstract}

Phase transitions are conventionally characterized by thermodynamic observables, but they also entail a reorganization of statistical structure. In finite systems, third-order pseudo-transitions reveal additional structural reorganization preceding or accompanying conventional criticality. Here we use the eigen-microstate approach to diagnose this reorganization from the full distribution of normalized spectral weights. Central moments about the uniform weight level characterize spectral concentration and asymmetric large-weight deviations, while spectral projection exposes residual organization after dominant modes are removed. Applications to Ising and Potts models on regular lattices and random regular networks reveal ordered-side modal redistribution and critical-window residual-weight reorganization that are not captured by leading-mode condensation alone. The resolvability of these features depends on the number of states, topology, and spectral truncation. High-temperature extrema in finite Wolff ensembles are sampling-protocol sensitive and are therefore not assigned equilibrium pseudo-transition temperatures. These results show that full and projected eigen-microstate spectra provide complementary diagnostics of higher-order structural organization.
\end{abstract}

%\keywords{Suggested keywords}%Use showkeys class option if keyword
                              %display desired
\maketitle

%\tableofcontents

\section{Introduction}

Phase transitions are traditionally characterized by nonanalytic behavior of thermodynamic observables in the thermodynamic limit, where singularities sharply distinguish ordered and disordered phases\cite{Landau,Goldenfeld,Cardy}.
In realistic finite systems, however, such singularities are replaced by smooth but structured reorganizations of statistical states.
Understanding how these structural changes emerge prior to or alongside conventional critical behavior remains an open problem in statistical physics and complex systems\cite{Binder1987,Privman1990,Dorogovtsev2008RMP,Scheffer2012,Rietkerk2021}.

In finite systems, higher-order pseudo-transitions provide a natural framework for these reorganizations.
Within the microcanonical ensemble\cite{Gross2001,Schnabel2011PRE},
they are identified from successive entropy derivatives through microcanonical inflection-point analysis (MIPA)\cite{Bachmann2018,sitarachu2022evidence}.
This framework offers a direct structural interpretation in terms of entropy geometry, but typically requires detailed knowledge of the density of states, which is difficult to obtain in large, networked, or nonequilibrium systems\cite{MartinMayor2007PRL,Fernandez2009PRE,Nogawa2011PRE,Schierz2016PRE,Rose2019PRE,Mozolenko2024PRECompare}.

In parallel, canonical approaches based on fluctuation theory provide experimentally accessible alternatives.
Cumulants of energy or order parameters encode higher-order anomalies without explicit density-of-states reconstruction\cite{binder1981finite,Challa1986}.
Meanwhile, data-driven methods based on principal component analysis and machine learning identify transition regions directly from configuration data, typically with an emphasis on feature extraction or dimensionality reduction rather than on the statistical organization of the full spectrum\cite{wang2017machine,hu2017discovering,Wang2016PRB,VanNieuwenburg2017,Carrasquilla2017NatPhys,Wetzel2017PRE,Rahaman2023,Muzzi2024}.
Within this line of work, the eigen-microstate framework assigns physical meaning to spectral modes by interpreting phase transitions as condensation phenomena in configuration space, characterized by the emergence of a dominant eigenmode with finite statistical weight\cite{Liu2022CPLRG,Wang2023,Hu2023PhysicaA,Liu2025EigenEntropy}.

Thermodynamic cumulants characterize higher-order fluctuations of preselected macroscopic observables, whereas existing eigen-microstate analyses also characterize leading-mode condensation and related spectral observables. We extend higher-order statistical analysis to the full weight distribution of the eigen-microstate covariance spectrum, thereby providing a complementary view of internal statistical structure rather than replacing conventional critical-point criteria. The spectral decomposition is constructed from the covariance operator in configuration space. Treating the normalized spectral weights as a probability distribution, we define the third-order ratio
\begin{equation}
R_3 = \frac{K_3}{(K_2)^3}
\end{equation}

which combines the third and second central moments to probe asymmetric spectral-weight redistribution beyond the overall concentration characterized by $K_2$. It is not the conventional standardized skewness and does not mathematically isolate an independent third-order component; rather, it provides a higher-order response to be examined together with its constituent moments.

Within this framework, extrema of $R_3$ are interpreted as spectral responses associated with higher-order modal reorganization. Projected responses remove leading modes from the normalized spectrum and test whether residual spectral layers retain identifiable structure that is not apparent in the full spectrum. This spectral-filtering procedure diagnoses statistical organization; it does not independently define a pseudo-transition.

We examine Ising and Potts models on regular lattices and random regular networks, covering both continuous and discontinuous transitions. The Ising model benchmarks ordered-side organization, critical-window residual-mode structure, and sampling sensitivity; the Potts and network calculations test how the number of states and topology affect spectral resolvability~\cite{Barghathi2014PRL,Schrauth2018PRE,CojaOghlan2023CMP}. Our aim is to establish where full-spectrum and projected diagnostics provide information beyond macroscopic observables and leading-mode condensation, not to outperform conventional thermodynamic measurements in locating $T_c$ or to propose a universal transition criterion.

\begin{figure*}[t]
\centering
\includegraphics[width=0.98\linewidth]{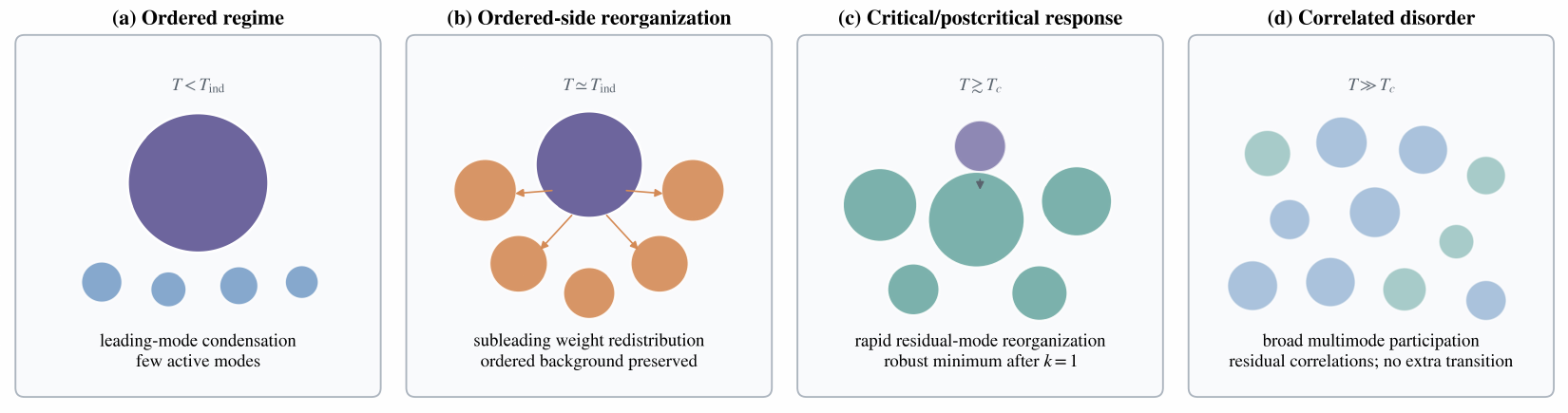}
\caption{Schematic spectral organization with increasing temperature: (a) leading-mode condensation, (b) ordered-side subleading redistribution near $T_{\mathrm{ind}}$, (c) critical/postcritical residual-mode reorganization after removing the first mode, and (d) correlated high-temperature disorder. No additional high-temperature transition is implied.}
\label{fig:organization_schematic}
\end{figure*}

Figure~\ref{fig:organization_schematic} summarizes the organizational sequence examined below: leading-mode condensation, ordered-side subleading redistribution, critical/postcritical residual-mode reorganization, and correlated high-temperature disorder.

\section{Eigen-Microstate Framework and Spectral Structural Response}

In this section, we formulate a spectral description of the statistical ensemble in configuration space. In this representation, conventional phase transitions correspond to condensation of a dominant mode, whereas higher-order pseudo-transitions are associated with subtler redistributions of spectral weight within the fluctuation subspace.

\subsection{Eigen-microstate representation of the statistical ensemble}

Following the eigen-microstate formalism developed in Ref.~\cite{Sun2021CTP}, we represent the statistical ensemble directly in configuration space.
This construction differs from standard PCA in that the spectrum is interpreted as a statistical measure of ensemble organization rather than a dimensionality-reduction tool.

For clarity, we briefly specify the models analyzed in this work. For the ferromagnetic Ising model,
\begin{equation}
\mathcal{H}_{\mathrm{Ising}}=-J\sum_{\langle ij\rangle}s_i s_j,
\qquad s_i=\pm 1,
\end{equation}
and for the $q$-state Potts model,
\begin{equation}
\mathcal{H}_{\mathrm{Potts}}=-J\sum_{\langle ij\rangle}\delta_{s_i,s_j},
\qquad s_i\in\{1,2,\dots,q\},
\end{equation}
where $J>0$ and $\langle ij\rangle$ denotes nearest-neighbor pairs (or network edges for random regular networks). We set $J=k_B=1$ throughout.

We collect $M$ statistically independent configurations at each fixed control parameter
(e.g., temperature) using Wolff-type cluster Monte Carlo updates\cite{Wolff1989PRL,Wolff1989PLB}.
Detailed sampling parameters are given in Appendix~\ref{app:sampling}.

We analyze two-dimensional regular lattices and random regular networks (RRNs). The RRNs are generated using the NetworkX function \texttt{random\_regular\_graph} with fixed degree four. Each undirected realization is held fixed throughout the temperature sweep, and the reported RRN results are averaged over independently generated realizations. For the regular lattices, the linear sizes are $L=64,72,84,100,$ and $128$, corresponding to $N=L^2=4096,5184,7056,10000,$ and $16384$ sites. For the RRNs, $N$ denotes the number of vertices: the Ising model is investigated for $N=5000,7000,8000,9000,$ and $10000$, and the Potts models for $N=4096,5184,7056,10000,$ and $16384$. Further simulation and sampling details are given in Appendix~\ref{app:sampling}.
Each microstate is described by a vector of $N$ degrees of freedom,
\begin{equation}
A^{(I)} =
\begin{pmatrix}
s^{(I)}_1 \\
s^{(I)}_2 \\
\vdots \\
s^{(I)}_N
\end{pmatrix},
\qquad I=1,\dots,M.
\end{equation}

Because we analyze both Ising and Potts systems, we construct the ensemble matrix from centered variables,
\begin{equation}
\tilde{s}_i^{(I)} = s_i^{(I)}-\bar{s}_i,
\qquad
\bar{s}_i=\frac{1}{M}\sum_{I=1}^{M}s_i^{(I)}.
\end{equation}
All sampled configurations are then assembled into the $N\times M$ ensemble matrix
\begin{equation}
\begin{aligned}
\mathbf{A} &= \big(\tilde{A}^{(1)},\tilde{A}^{(2)},\dots,\tilde{A}^{(M)}\big), \\
\tilde{A}^{(I)} &= (\tilde{s}_1^{(I)},\tilde{s}_2^{(I)},\dots,\tilde{s}_N^{(I)})^T.
\end{aligned}
\end{equation}

The ensemble matrix is globally normalized to ensure consistent statistical weights,
\begin{equation}
\mathbf{A} \rightarrow 
\frac{\mathbf{A}}{\sqrt{\sum_{I=1}^{M}\sum_{i=1}^{N}(\tilde{s}_i^{(I)})^2}}.
\end{equation}
At each control-parameter value, this equation multiplies every matrix element by the same nonzero scalar $c$. The eigenvalues therefore transform as $\lambda_\alpha\to c^2\lambda_\alpha$, leaving the normalized weights $w_\alpha$, and hence $R_3$, unchanged. A change in the shape of an $R_3$ curve would require an additional operation, such as altered centering, relative rescaling, or spectral selection, rather than this global rescaling alone.

We perform singular value decomposition,
\begin{equation}
\mathbf{A} = \mathbf{U}\boldsymbol{\Sigma}\mathbf{V}^{T},
\end{equation}
where $\boldsymbol{\Sigma}=\mathrm{diag}(\sigma_1,\sigma_2,\dots)$ with 
$\sigma_1\ge\sigma_2\ge\dots\ge0$; here the columns of $\mathbf{V}$ encode how each spectral mode is represented across the sampled microstates, i.e., the ensemble-evolution information in sample space, but that aspect is not the focus of the present work.

The reduced singular value decomposition is performed directly on the rectangular ensemble matrix, without explicitly constructing an $N\times N$ correlation matrix.

The columns of $\mathbf{U}$ define orthonormal eigen microstates in the configuration space,
\begin{equation}
E^{(\alpha)} = \mathbf{U}_{:,\alpha},
\end{equation}
where $E^{(\alpha)}$ denotes the $\alpha$th eigen microstate, namely the $\alpha$th column of $\mathbf{U}$, whose components give the spatial structure of that mode in configuration space.

The configuration-space covariance operator is
\begin{equation}
\mathbf{C}=\mathbf{A}\mathbf{A}^{T}
=\mathbf{U}\boldsymbol{\Sigma}^{2}\mathbf{U}^{T},
\qquad \lambda_\alpha=\sigma_\alpha^{2}.
\end{equation}
Thus the columns of $\mathbf{U}$ are eigenvectors of $\mathbf{C}$ and define the configuration-space eigen microstates. By contrast, $\mathbf{A}^{T}\mathbf{A}=\mathbf{V}\boldsymbol{\Sigma}^{2}\mathbf{V}^{T}$ acts in sample space; its eigenvectors are the columns of $\mathbf{V}$, not configuration-space eigen microstates. The two operators have the same nonzero eigenvalues, so this notational correction does not alter the nonzero spectrum or any numerical result.
The spectral decomposition used throughout this work is therefore defined with respect to the configuration-space covariance operator $\mathbf{C}$, whose eigenbasis provides an intrinsic representation of ensemble fluctuations.

We define normalized spectral weights
\begin{equation}
w_\alpha=\frac{\lambda_\alpha}{\sum_\beta\lambda_\beta},
\qquad
\sum_\alpha w_\alpha=1.
\end{equation}
These weights are not empirical ranking parameters; rather, they are fluctuation weights induced by the eigenvalue spectrum of the configuration-space covariance operator $\mathbf{C}$. The distribution $\{w_\alpha\}$ characterizes the geometric organization of the ensemble. All spectral moments and response ratios below are evaluated from these normalized spectral weights, not directly from the unnormalized eigenvalues.

\subsection{Eigen-microstate condensation and phase transitions}

Within this framework, phase transitions acquire a direct geometric interpretation.

In the disordered regime, spectral weights are broadly distributed and 
\begin{equation}
\lim_{M\to\infty} w_\alpha = 0
\end{equation}
for all $\alpha$.

In contrast, when the largest weight satisfies
\begin{equation}
\lim_{M\to\infty} w_1 > 0,
\end{equation}
the leading eigen microstate condenses in ensemble space.

This condensation is mathematically analogous to Bose--Einstein condensation 
and physically corresponds to the emergence of macroscopic structural order. 
In conventional second-order phase transitions, the ordered phase is 
characterized by such eigen-microstate condensation.

\begin{center}
\textit{Eigen-microstate condensation therefore provides a geometric view of phase transitions.}
\end{center}

This interpretation does not rely on predefined order parameters; 
instead, macroscopic ordering is identified through spectral localization 
in configuration space.

\subsection{Spectral generalized response for third-order structural reorganization}

Leading-mode condensation captures ordinary ordering, but not all finite-size reorganization. Additional changes may occur within the fluctuation subspace, where no single mode becomes macroscopic while spectral weight is redistributed among subleading modes.

We therefore analyze the normalized weight distribution $\{w_\alpha\}$ through central moments defined relative to the uniform level $\bar{w}=1/M$,
\begin{equation}
K_n = \sum_{\alpha} (w_\alpha - \bar{w})^n.
\end{equation}
Here $K_2$ is the second central moment and measures overall concentration of spectral weight. The third central moment $K_3$ gives greater weight to large deviations from the uniform level, especially modes with large positive $w_\alpha-\bar w$, and therefore probes asymmetric redistribution across modes. It is the lowest-order central moment sensitive to asymmetry of the spectral-weight distribution.

Reference~\cite{Liu2026arXiv} introduces the energy-cumulant ratio $\Xi=\kappa_3/(\kappa_2)^3$. We borrow its cubic denominator as an empirical convention for the distinct spectral moments $K_2$ and $K_3$; this is not a derivation of a spectral canonical criterion, and $(K_2)^3$ is not a unique or universal normalization. Because $K_3$ alone is strongly influenced by the second-order concentration scale, we define
\begin{equation}
R_3 = \frac{K_3}{(K_2)^3},
\end{equation}
which compares the third-order response with the second-order concentration scale. This is not the conventional standardized skewness, for which the denominator would be $(K_2)^{3/2}$, and $R_3$ does not independently isolate third-order spectral asymmetry. Instead, it provides a single higher-order response whose extrema can be traced back to $K_2$ and $K_3$; Supplemental Fig.~\ref{fig:ising_critical_moments} presents these quantities together to make that relationship explicit.

The ratio $R_3$ is not a microcanonical entropy derivative and does not independently define a third-order pseudo-transition. The mapping from thermodynamic fluctuations to covariance spectra is nonlinear, as illustrated by the two-state example in Appendix~\ref{app:nonlinear}. Extrema of $R_3$ therefore define spectral anomalies whose correspondence with geometric or microcanonical reorganization must be established using independent indicators.

\subsection{Projected spectral response as spectral tomography}

To examine how spectral reorganization is distributed across different modal layers, we define a projected spectral distribution by removing the leading $k$ modes,
\begin{equation}
\tilde{w}^{(k)}_{\alpha}
=
\frac{w_{\alpha+k}}{1-\sum_{i=1}^{k}w_i}.
\end{equation}
The corresponding central moments and response ratio are constructed in the same way as for the full spectrum. We denote them by $\tilde K_n^{(k)}$ and $\tilde R_3^{(k)}=\tilde K_3^{(k)}/[\tilde K_2^{(k)}]^3$, respectively; the uniform reference level is evaluated for the remaining $M-k$ weights.

This projection is a spectral-filtering procedure: increasing $k$ exposes successively deeper residual spectra and reveals whether an anomaly extends beyond the leading ordering channel. There is no universal truncation depth, and the evolution with $k$ characterizes spectral-layer organization rather than defining an independent transition temperature.

Full-spectrum and projected extrema are interpreted as modal counterparts of structural reorganization when they coincide with independent geometric or microcanonical indicators. Weak, shifted, or absent projected anomalies instead indicate limited spectral resolvability, not necessarily the absence of geometric restructuring. The main text emphasizes full-spectrum and $k=1$ responses; the Ising benchmark and Supplemental Material provide deeper truncations and robustness checks.

\subsection{Spectral participation and high-temperature limit}
The spectral participation of eigen microstates can be quantified by the effective number of active modes,
\begin{equation}
R_{\mathrm{eff}} = \frac{1}{\sum_\alpha w_\alpha^2},
\end{equation}
which measures how many modes contribute significantly to the ensemble. 

In the ideal high-temperature limit of completely random and uncorrelated configurations, the ensemble matrix approaches a random matrix with independent entries. In this regime, the spectral statistics are well approximated by Wishart-type random matrices\cite{Marchenko1967MatSb}, 
leading to a theoretical estimate for the participation ratio,
\begin{equation}
R_{\mathrm{eff}}(T \to \infty) \approx \frac{MN}{M+N},
\label{eq:Reff}
\end{equation}
where $N$ is the number of degrees of freedom and $M$ is the number of sampled configurations. This estimate serves only as a reference benchmark for spectral participation at fixed $(M,N)$.

If residual correlations persist, mode competition reduces the number of effectively active independent directions, and one expects
\begin{equation}
R_{\mathrm{eff}} < \frac{MN}{M+N}.
\end{equation}

Therefore, deviations below this theoretical limit provide a direct indicator of nontrivial correlations and remaining structural organization in the ensemble. As shown in the Results section, our data do not fully saturate the random benchmark, indicating that the high-temperature regime is fluctuation-dominated but not perfectly uncorrelated.

\section{Results}
\subsection{Benchmark: 2D Ising model}

\begin{figure*}[t]
\centering
\includegraphics[width=0.85\linewidth]{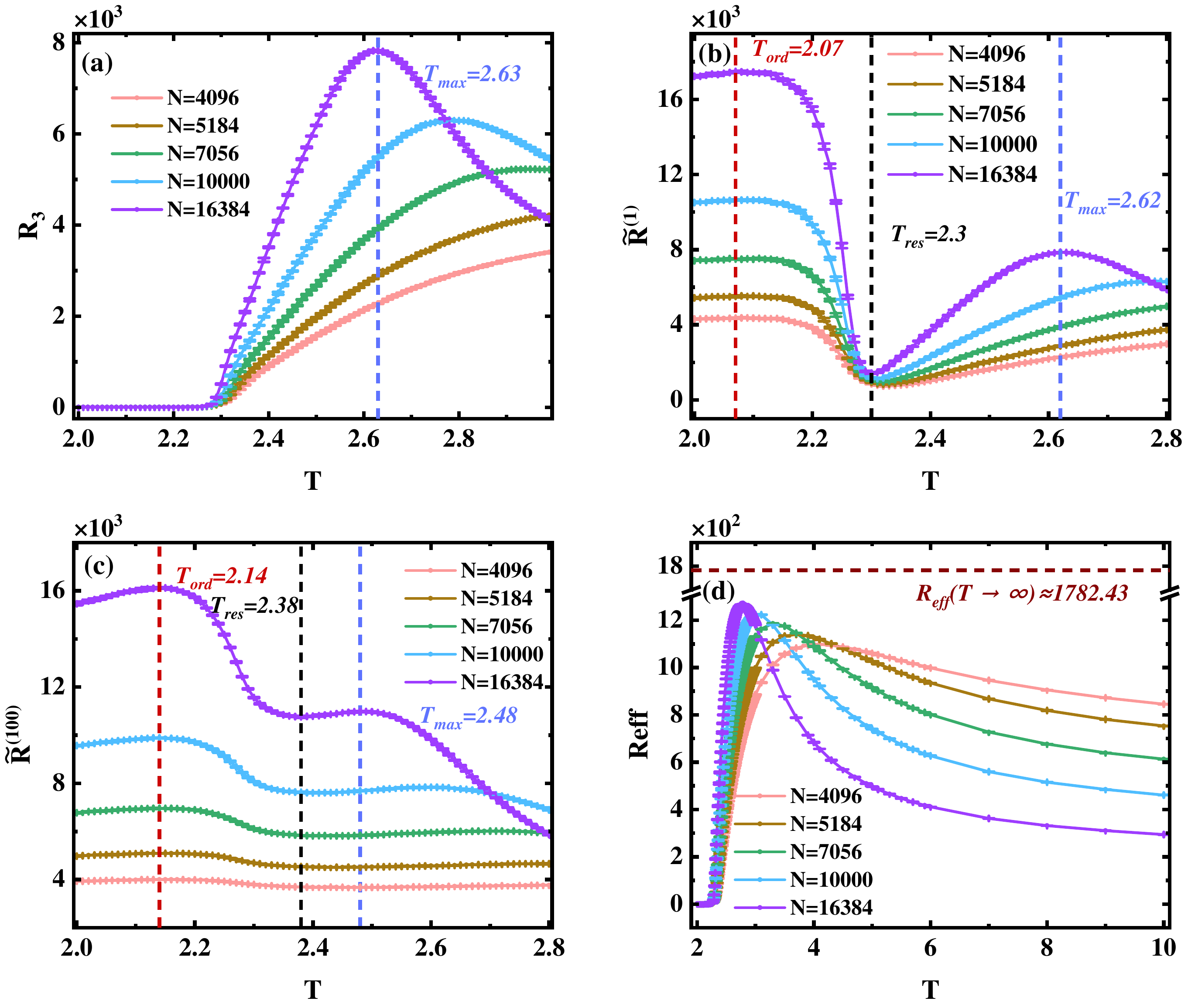}
\caption{
Spectral responses of the two-dimensional Ising model.
(a)~Full-spectrum response~$R_3$.
(b)~Projected response~$\tilde{R}_3^{(1)}$.
(c)~Projected response~$\tilde{R}_3^{(100)}$.
(d)~Effective spectral dimension~$R_{\mathrm{eff}}$.
Red, black, and blue dashed lines mark ordered-side responses ($T_{\mathrm{ord}}$), critical/postcritical residual responses ($T_{\mathrm{res}}$), and empirical high-temperature maxima of finite Wolff spectra ($T_{\max}$), respectively. The last are not equilibrium transition assignments.
The horizontal dashed line in panel~(d) marks the random-matrix benchmark for the largest system size, $N=16384$.
}
\label{fig:ising_benchmark}
\end{figure*}

The two-dimensional Ising model provides a natural benchmark because its ordinary ordering transition is captured by leading-mode condensation~\cite{Hu2026CPL}. Figure~\ref{fig:ising_benchmark} summarizes the full-spectrum, projected, and participation responses used here to examine organization beyond that dominant mode.

On the ordered side, removing the leading mode exposes a weak but recurrent response near $T_{\mathrm{ind}}$. This feature remains visible under further truncation, although its position and strength depend on the truncation depth. Its recurrence in independent spectra and its proximity to independently identified structural reorganization support the interpretation that the third-order reorganization is accompanied by asymmetric redistribution of subleading spectral weight. It is a complementary spectral characterization of ordered-side reorganization, not a new critical temperature or an independently established pseudo-transition.

\begin{figure*}[t]
\centering
\includegraphics[width=0.85\linewidth]{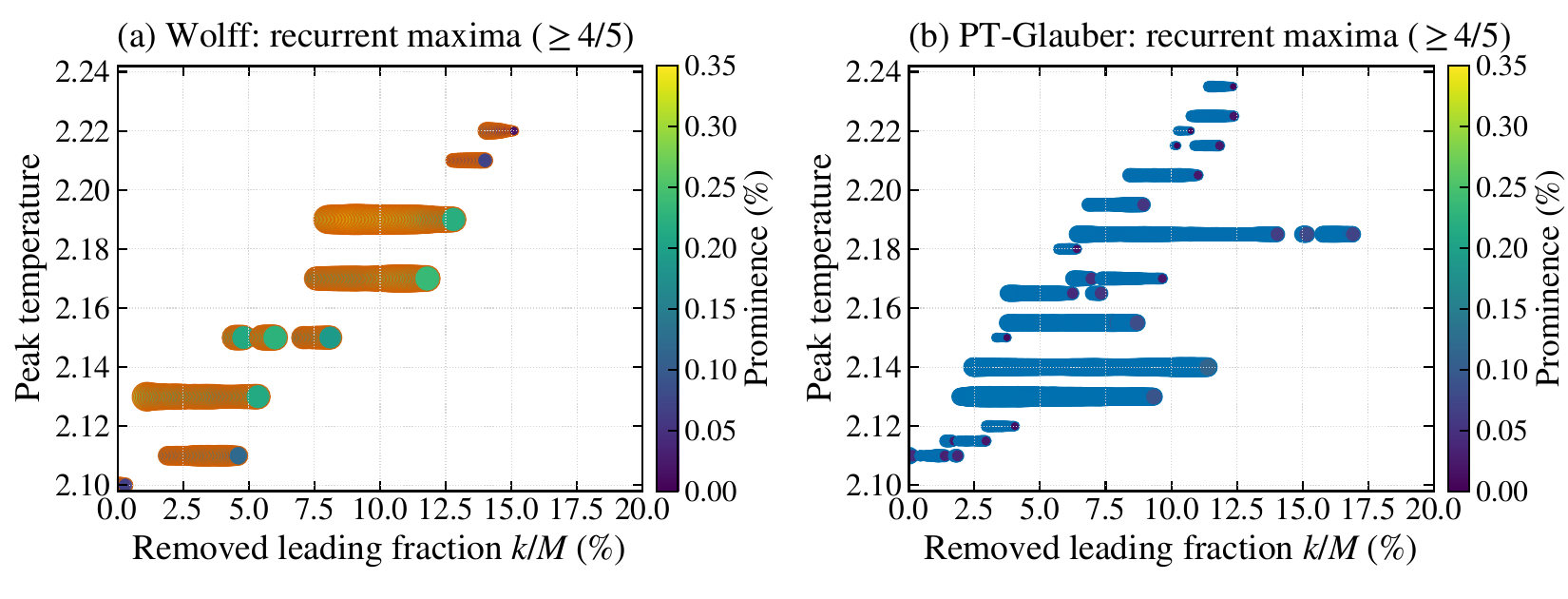}
\caption{Recurrent spectral-response maxima for Wolff (a) and PT-Glauber (b) spectra at $L=128$. Maxima recurring in at least four of five independent spectra are shown as functions of the removed leading fraction $k/M$.}
\label{fig:ising_truncation_fraction}
\end{figure*}

A denser truncation analysis is shown in Fig.~\ref{fig:ising_truncation_fraction}, where the truncation depth is expressed as the fraction $f=k/M$. For the Wolff spectra, successive temperature ridges appear as leading spectral components are removed, showing that different spectral layers expose weak nonmonotonic organization at different temperatures. Within the common window $2.10\leq T\leq2.24$, recurrent Wolff maxima persist up to $f\simeq0.151$, while no strict local maximum is found at $f=0.20$. The PT-Glauber analysis shows the same qualitative spectral-layer exposure and eventual disappearance.

% Sampling cross-check moved to Supplemental Fig. S11.

At high temperature, local maxima of $R_3$ appear in spectra constructed from finite Wolff trajectories. In the $L=128$ cross-check over the temperatures and full/shallow projections shown in Fig.~\ref{fig:ising_sampling_check_supp}, they are not reproduced as robust extrema by PT-Glauber sampling. We therefore treat them as sampling-protocol-sensitive features within that tested range and do not assign them equilibrium pseudo-transition temperatures.

A distinct feature occurs in the critical window. The full-spectrum $R_3$ has no local minimum between $T=2.20$ and $2.40$. After removal of the leading mode, however, both Wolff and PT-Glauber spectra exhibit a local minimum of $\tilde R_3^{(1)}$ at $T=2.300$ (Fig.~\ref{fig:ising_sampling_check_supp}). The spectral moments in Fig.~\ref{fig:ising_critical_moments} show that the leading remaining weight $\tilde w_2^{(1)}$, $\tilde K_2^{(1)}$, and $\tilde K_3^{(1)}$ peak in the same critical/postcritical region while their ratio reaches a minimum. These results identify rapid residual-weight reorganization accompanying the breakdown of leading-mode condensation. Within the tested window, this is a sampling-robust spectral feature, not an independently established transition temperature.

As shown in Fig.~\ref{fig:ising_benchmark}(d), $R_{\mathrm{eff}}$ remains below the random-matrix benchmark of Eq.~(\ref{eq:Reff}), indicating residual correlations in the high-temperature configuration spectrum. The leading eigen microstates in Fig.~\ref{fig:ising_U} likewise evolve from a nearly uniform ordered pattern to fragmented high-temperature structures.

\begin{table}[t]
\caption{Ordered-side spectral and structural indicators for the two-dimensional Ising model. The spectral range corresponds to representative projection depths rather than statistical error.}
\label{tab:ising_compare}
\begin{ruledtabular}
\begin{tabular}{lc}
Observable & Ordered-side characteristic $T$ \\
\hline
Spectral response & $2.08$--$2.14$ \\
Isolated-spin density & 2.229 \\
Microcanonical (MIPA) & 2.230 \\
\end{tabular}
\end{ruledtabular}
\end{table}

Table~\ref{tab:ising_compare} compares the ordered-side spectral feature with independent structural indicators. The ordered-side $R_3$ response provides a spectral representation of subleading modal redistribution within the ordered background.

\subsection{Applicability across models and topologies}

For the random-regular-network and Potts analyses presented below, high-temperature extrema obtained from Wolff ensembles are regarded as empirical spectral features under the stated sampling protocol. Unless supported by an explicit sampling cross-check, they are not assigned equilibrium dependent-pseudo-transition temperatures. The main emphasis is therefore on the ordered-side response and on the resolvability of spectral organization across models and topologies.

We next examine the applicability of the spectral response across topology and model class, focusing on the Ising model on random regular networks and the Potts model on both regular lattices and random regular networks.

\subsubsection{Ising model on random regular networks}

To test how the spectral reorganization is affected by topological disorder, we next consider the Ising model on random regular networks (RRNs)~\cite{Bollobas2001RandomGraphs}. Unlike the two-dimensional regular lattice, an RRN has fixed degree but lacks both translational symmetry and Euclidean geometric embedding. It therefore provides a useful setting for examining whether the modal reorganizations observed on the regular lattice remain visible when the underlying spatial structure is removed.

Figure~\ref{fig:RRNisingW} shows that the size evolution of $W_2/W_1$, the leading spectral weights, and the finite-size scaling of $W_1$ consistently capture the ordinary ordering transition on RRNs~\cite{Sun2021CTP}. These quantities establish the principal ordering channel but do not resolve higher-order redistribution; we therefore examine the full-spectrum and projected responses to determine whether subleading reorganization remains visible without spatial embedding and translational symmetry.

\begin{figure*}[t]
\centering
% panels
\includegraphics[width=0.85\linewidth]{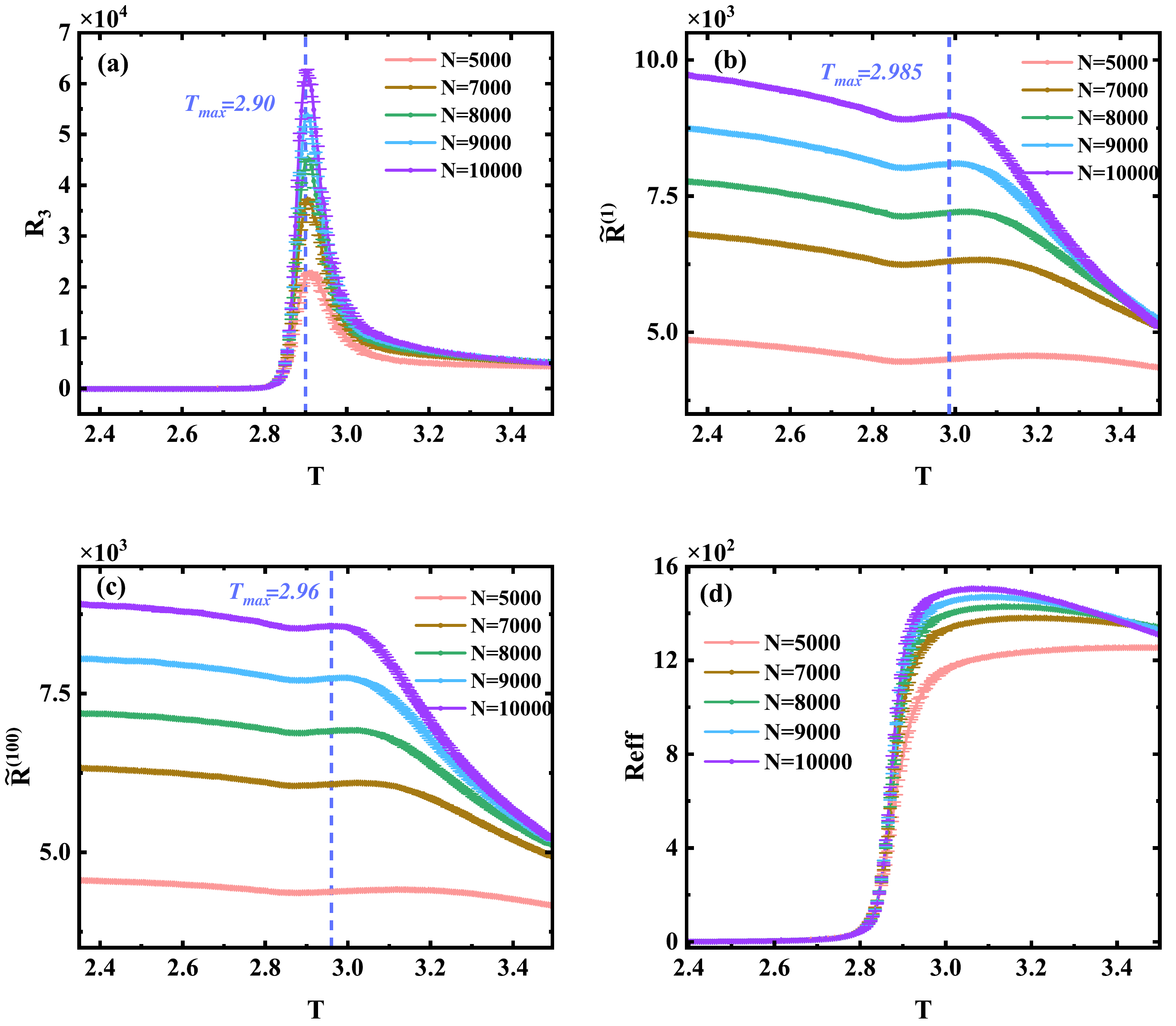}
\caption{
Spectral responses of the Ising model on random regular networks.
(a)~Full-spectrum response~$R_3$.
(b)~Projected response~$\tilde{R}_3^{(1)}$.
(c)~Projected response~$\tilde{R}_3^{(100)}$.
(d)~Effective spectral dimension~$R_{\mathrm{eff}}$.
Blue dashed lines mark empirical maxima of finite Wolff spectra, not geometric or microcanonical transition indicators.
Panels~(a)--(c) illustrate the truncation-depth dependence of modal reorganization on random regular networks.
}
\label{fig:RRNisingResult}
\end{figure*}

Figure~\ref{fig:RRNisingResult} shows the full-spectrum and projected spectral responses for the Ising model on random regular networks. In Fig.~\ref{fig:RRNisingResult}(a), the finite-Wolff ensemble exhibits a sharp high-temperature maximum. In the absence of a corresponding sampling-convergence check for these network ensembles, we retain this feature only as an empirical spectral anomaly under the stated sampling protocol and do not assign it a dependent pseudo-transition temperature.

After removing the leading mode, Fig.~\ref{fig:RRNisingResult}(b) shows that this empirical feature is strongly reduced and shifted, while further truncation in Fig.~\ref{fig:RRNisingResult}(c) produces a broader and less sharply localized residual response. This behavior demonstrates sensitivity to the leading spectral sector and to the truncation depth, but it does not by itself establish an equilibrium disordered-side structural mechanism.

Figure~\ref{fig:RRNisingU} clarifies the reduced resolvability: the leading mode captures global ordering, whereas the second mode appears as scattered local hotspots rather than a coherent mesoscopic domain. Topological disorder thus fragments the subleading fluctuation sector into weak, spatially incoherent components, producing broader and more truncation-sensitive projected responses than on the regular lattice.

Figure~\ref{fig:RRNisingResult}(d) provides the corresponding spectral-participation background. The effective spectral dimension $R_{\mathrm{eff}}$ remains small at low temperature, reflecting concentration in a few leading modes, and then rises sharply across the transition region. This behavior indicates enhanced but still correlated modal participation; it does not establish a separate equilibrium high-temperature pseudo-transition.

\subsubsection{Potts models: effects of the number of states and topology}

We next turn to the Potts model to examine how spectral reorganization depends on both the number of states and the underlying topology. 
We compare the cases $q=3$ and $q=8$ on regular lattices and random regular networks. 
Potts models on complex networks have been widely studied, and their transition behavior is known to depend sensitively on the underlying topology and degree distribution~\cite{Dorogovtsev2004EPJB}.

\begin{figure*}[p]
\centering
\includegraphics[width=0.85\linewidth,height=0.40\textheight,keepaspectratio]{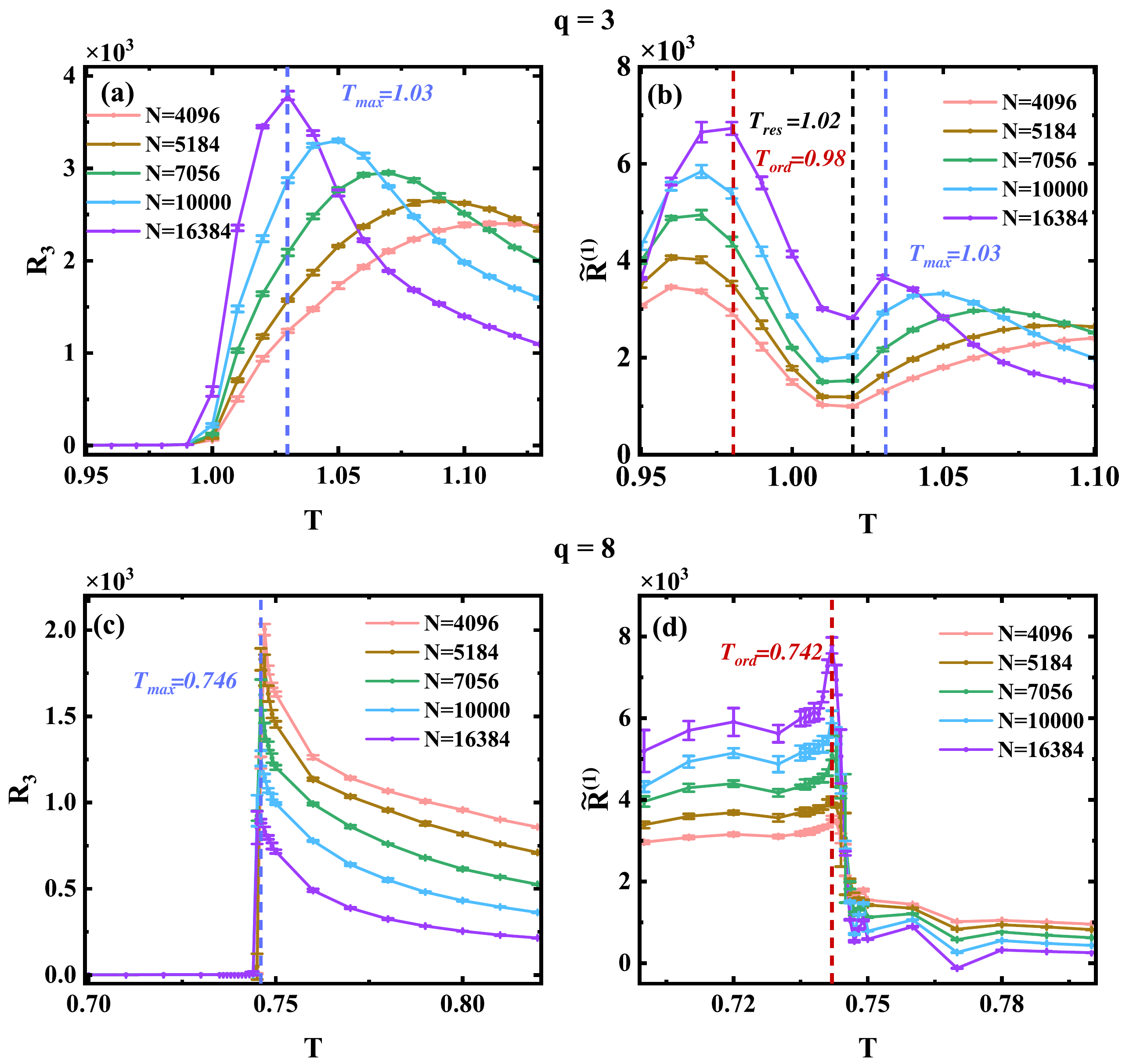}
\caption{Regular-lattice Potts responses for $q=3$ [(a),(b)] and $q=8$ [(c),(d)]: full-spectrum $R_3$ [(a),(c)] and $k=1$ projected response [(b),(d)]. Blue dashed lines mark empirical Wolff maxima, red lines mark ordered-side response regions, and the black line marks the projected residual-response minimum. Complete responses are given in Figs.~\ref{fig:Q3Potts_regular} and \ref{fig:Q8Potts_regular}.}
\label{fig:Potts_regular_compact}
\end{figure*}
\begin{figure*}[p]
\centering
\includegraphics[width=0.85\linewidth,height=0.40\textheight,keepaspectratio]{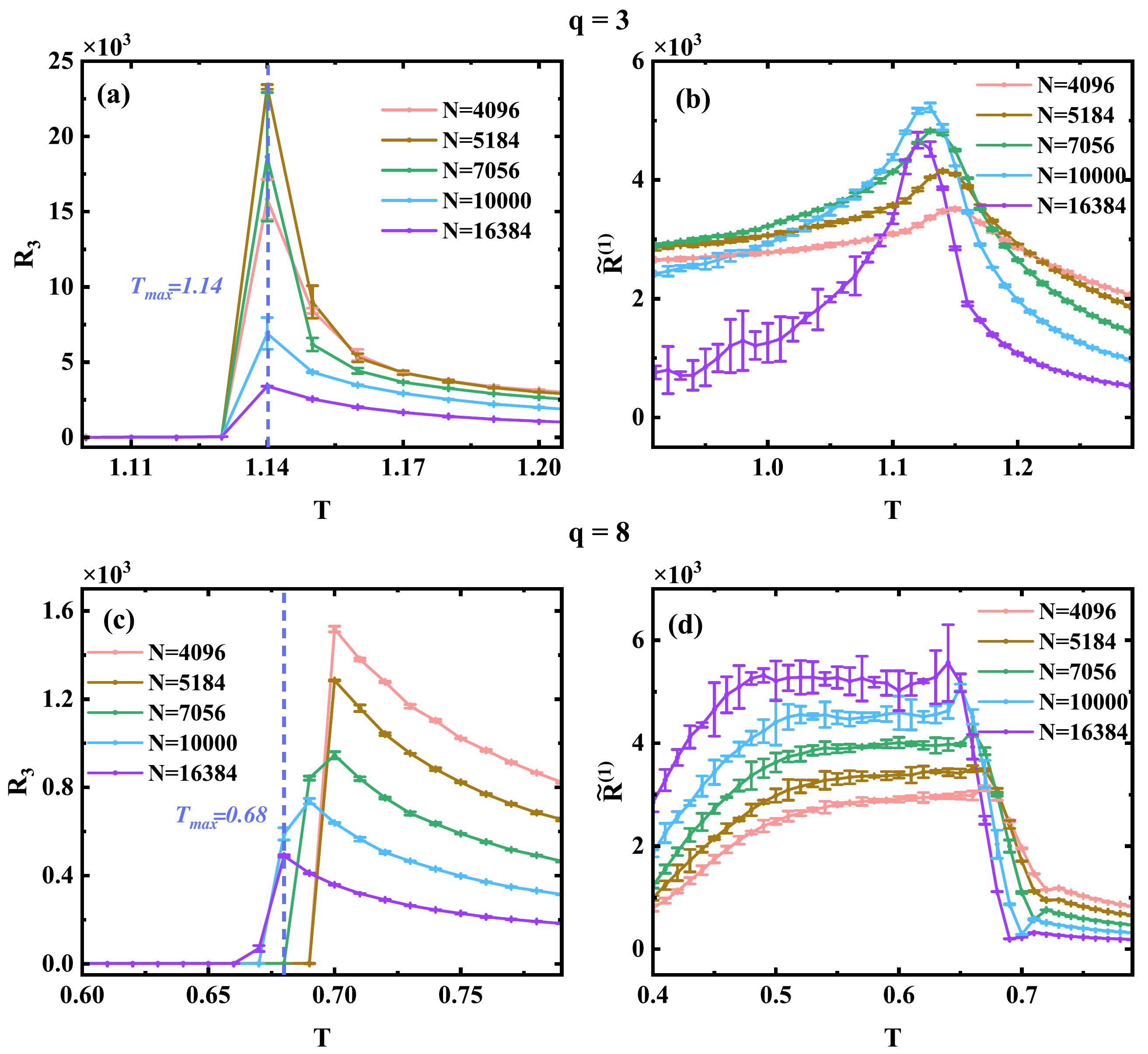}
\caption{RRN Potts responses for $q=3$ [(a),(b)] and $q=8$ [(c),(d)]: full-spectrum $R_3$ [(a),(c)] and $k=1$ projected response [(b),(d)]. Blue dashed lines mark nominal empirical extrema of averaged finite-Wolff curves; they do not imply cleanly separated anomalies. Complete responses are given in Figs.~\ref{fig:Q3Potts_rrn} and \ref{fig:Q8Potts_rrn}.}
\label{fig:Potts_RRN_compact}
\end{figure*}

\begin{table*}[t]
\caption{Characteristic spectral-feature temperatures for regular-lattice Ising and for Potts models on regular lattices and RRNs.}
\label{tab:third_order_all_models}
\centering
\begin{ruledtabular}
\begin{tabular}{lcccccc}
Model & Topology & $N=4096$ & $N=5184$ & $N=7056$ & $N=10000$ & $N=16384$ \\
\hline
2D Ising & regular lattice & 2.14 & 2.16 & 2.17 & 2.19 & $2.08$--$2.14$ \\
\hline
\multirow{2}{*}{Potts $q=3$}
& regular lattice & 0.97 & 0.97 & 0.98 & 0.98 & 0.98 \\
& RRN & 1.110 & 1.120 & 1.120 & 1.130 & 1.130 \\
\hline
\multirow{2}{*}{Potts $q=8$}
& regular lattice & 0.743 & 0.743 & 0.744 & 0.744 & 0.742 \\
& RRN & 0.69 & 0.68 & 0.68 & 0.67 & 0.66 \\
\end{tabular}
\end{ruledtabular}
\end{table*}

The RRN Potts entries in Table~\ref{tab:third_order_all_models} are nominal empirical extrema of the averaged finite-Wolff curves. They are not equilibrium transition assignments and do not imply that the corresponding anomalies are cleanly isolated.

As indicated by the leading-spectrum condensation results in Figure~\ref{fig:PottsW}, the ordinary transition of the eight-state Potts model remains well captured by eigen-microstate condensation. The finite-size fits on the two sides of the main-transition region show that the effective exponent is nearly zero below the transition but becomes positive above it, supporting the first-order nature of the transition. These leading-spectrum quantities characterize the principal ordering channel, whereas the full-spectrum and projected responses are used below to examine how higher-order modal reorganization is distributed across the residual spectrum.

For the regular-lattice Potts model with $q=3$, Fig.~\ref{fig:Potts_regular_compact}(a) exhibits the empirical high-temperature Wolff maximum described above. The $k=1$ response in Fig.~\ref{fig:Potts_regular_compact}(b) reveals an ordered-side feature; Fig.~\ref{fig:Q3Potts_regular} shows its persistence at $k=100$ and the corresponding $R_{\mathrm{eff}}$ background. This projected response provides the main evidence for subleading spectral organization.

For the regular-lattice Potts model with $q=8$, Figs.~\ref{fig:Potts_regular_compact}(c) and \ref{fig:Potts_regular_compact}(d) show that the full and projected responses are compressed near the main-transition window around $T_c=0.746$. The deeper projection and $R_{\mathrm{eff}}$ in Fig.~\ref{fig:Q8Potts_regular} are likewise concentrated in this window, and no distinct high-temperature spectral branch is resolved. Thus, increasing $q$ compresses the resolvable spectral reorganization into the main-transition region.

On random regular networks, the three-state Potts model provides a boundary case for spectral filtering. Figure~\ref{fig:Potts_RRN_compact}(a) retains the empirical high-temperature Wolff maximum, whereas the $k=1$ response in Fig.~\ref{fig:Potts_RRN_compact}(b) does not cleanly isolate an ordered-side anomaly. The complete projections in Fig.~\ref{fig:Q3Potts_rrn} are broadened and truncation-sensitive, indicating structural reorganization fragmented over the residual spectrum.

For the eight-state Potts model on random regular networks, Figs.~\ref{fig:Potts_RRN_compact}(c) and \ref{fig:Potts_RRN_compact}(d) show a further reduction of spectral resolvability, with the response concentrated near the main-transition region. The deeper projection and $R_{\mathrm{eff}}$ in Fig.~\ref{fig:Q8Potts_rrn} are similarly confined to this window. The combined effects of larger $q$ and topological disorder therefore compress the resolvable response; they do not establish a separate high-temperature equilibrium mechanism.

Figure~\ref{fig:potts_U} provides a complementary configuration-space view of the dominant mode for the eight-state Potts model on a regular lattice. At $T=0.71$, the dominant eigen microstate is nearly uniform, indicating a coherent ordered pattern. Across the main-transition window, this background is rapidly disrupted, and the dominant mode becomes markedly more heterogeneous and fragmented at $T=0.75$ and $T=0.85$. This behavior is consistent with the spectral analysis in Fig.~\ref{fig:Potts_regular_compact}(c,d), where the relevant reorganization is concentrated near the main transition rather than separated into a distinct high-temperature feature.

\section{Discussion}

The central contribution of $R_3$ is to extend higher-order statistical analysis from preselected macroscopic observables to the full eigen-microstate weight distribution. The second and third central moments separately quantify concentration and asymmetric large-weight deviations, while their ratio supplies a higher-order response to spectral-weight redistribution. This complements, but does not replace, thermodynamic cumulants and leading-mode condensation. Geometric or microcanonical indicators remain necessary to establish the physical correspondence of a spectral response.

Full and projected spectra need not exhibit the same extrema because projection removes the dominant ordering channel and renormalizes the remaining weights. The Ising ordered-side response near $T_{\mathrm{ind}}$ illustrates recurrent subleading organization, while the $k=1$ minimum at $T=2.300$ demonstrates critical/postcritical residual-weight reorganization in both tested samplers. Neither projected feature is an independent transition criterion. Across the Potts and network examples, increasing $q$ and introducing topological disorder can compress or fragment the resolvable response; broad, shifted, or absent projected anomalies therefore indicate limited spectral resolvability rather than the absence of geometric restructuring.

Sampling robustness is a separate requirement from spectral resolvability. The high-temperature Wolff extrema are not correspondingly reproduced in the tested $L=128$ PT-Glauber full and shallow-projected spectra and must therefore remain sampling-protocol-sensitive features. This comparison neither establishes a mechanism for those extrema nor licenses a general equilibrium interpretation outside the tested temperatures and truncation depths.

The method is intended as a structural diagnostic, not as a computationally cheaper replacement for conventional critical-point measurements. For a dense reduced SVD of the $N\times M$ ensemble matrix with $N\gg M$, a standard leading-work estimate is $O(NM^2)$, with lower-order $O(M^3)$ work, and storing the matrix requires $O(NM)$ entries. Explicit construction of an $N\times N$ covariance matrix is unnecessary. This estimate does not describe a truncated solver that computes only selected singular modes, nor does it assert a runtime advantage. This is more demanding than measuring standard thermodynamic observables, but it provides layer-resolved information about internal modal organization. Further work is needed to test finite-size convergence, sampler independence, and truncation robustness in other systems.

\section{Conclusion}

We introduced a higher-order statistical diagnostic based on normalized eigen-microstate spectral weights. The ratio $R_3$ combines third- and second-order central moments to probe spectral-weight redistribution, while projected responses expose structure in residual modal layers. In the Ising benchmark, recurrent ordered-side organization near $T_{\mathrm{ind}}$ and the sampling-robust $k=1$ minimum at $T=2.300$ demonstrate complementary information beyond leading-mode condensation. The Potts and random-regular-network examples show how the number of states and topology can compress or fragment the resolvable response.

The framework therefore supplements conventional thermodynamic and leading-spectrum analyses; it is not a universal transition criterion. High-temperature Wolff extrema remain sampling-protocol-sensitive where corresponding PT-Glauber checks do not reproduce them, and projected features can depend on finite size and truncation depth. Independent structural indicators and sampling checks remain essential for physical interpretation. Future work should test finite-size convergence, sampler independence, and truncation robustness in experimentally sampled configurations, nonequilibrium systems, and other many-body settings.

\begin{acknowledgments}
This work is supported by the National Natural Science Foundation of China (Grant No. 12575033 and 12304257).
\end{acknowledgments}

\appendix

\section{Sampling details}
\label{app:sampling}

For the Ising model, we discard the first $4\times10^5$ Monte Carlo steps for equilibration and then sample configurations every $800$ steps over the following $1.6\times10^6$ steps. For the Potts model, we discard the first $8\times10^4$ steps and then sample configurations every $400$ steps over the following $4\times10^5$ steps. These sampling windows yield $M=2000$ configurations per temperature for the Ising model and $M=1000$ for the Potts models. For the largest regular-lattice systems, the corresponding ensemble matrices have dimensions $16384\times2000$ and $16384\times1000$, respectively. The RRN results are averaged over $20$ independently generated network realizations, with each topology held fixed throughout the temperature sweep. No additional smoothing or interpolation is applied to the spectral observables. The temperature step is $\Delta T=0.005$ near the main transitions, $\Delta T=0.01$ in nearby regions, and $\Delta T=0.1$ well above the regime where the disordered-side spectral anomaly may occur. For the $L=128$ regular-lattice Ising analysis, additional independent Wolff spectra and PT-Glauber cross-checks are used to test sampling robustness.

\section{Nonlinear mapping between thermodynamic and spectral descriptions}
\label{app:nonlinear}

The relation between thermodynamic fluctuations and spectral weights is intrinsically nonlinear, even in the simplest coexistence scenario. Consider two competing macroscopic configurations, $A$ and $B$, with canonical probabilities $P_A$ and $P_B$. In configuration space, the ensemble may then be viewed heuristically as a mixture of two Gaussian clusters centered at $\mathbf{m}_A$ and $\mathbf{m}_B$. The covariance matrix contains both intra-state fluctuations and an inter-state contribution proportional to $P_A P_B$, so the leading eigenvalue scales as
\begin{equation}
\lambda_1 \propto P_A P_B \lVert \mathbf{m}_A-\mathbf{m}_B \rVert^2,
\end{equation}
while the remaining eigenvalues provide a background contribution $\Lambda_{\mathrm{bg}}$. The associated spectral weight is therefore
\begin{equation}
w_1 = \frac{\lambda_1}{\lambda_1 + \Lambda_{\mathrm{bg}}},
\end{equation}
which depends nonlinearly on $P_A P_B$ and exhibits saturation rather than simple proportionality. Even in this minimal two-state setting, the spectral weights are nonlinear functions of the underlying thermodynamic probabilities. Accordingly, $R_3$ should be interpreted as encoding redistribution of statistical weight only indirectly, rather than as an explicit derivative-based entropy diagnostic.

\bibliography{apssamp}% Produces the bibliography via BibTeX.

% Start the supplemental material on a new, single-column page.
\onecolumngrid
\clearpage
% Included by Final_Manuscript_clean.tex after the references.
\section*{Supplemental Material}
% To show revision additions in blue, replace #1 with \textcolor{blue}{#1}.
\newcommand{\newrev}[1]{#1}

\renewcommand{\thefigure}{S\arabic{figure}}
\setcounter{figure}{0}

\newrev{Throughout the Supplemental Material, $K_n$ denotes the $n$th central moment of normalized spectral weights about the uniform level, $R_3=K_3/(K_2)^3$, and $\tilde R_3^{(k)}$ is the corresponding response after removing $k$ leading modes. The effective spectral dimension is $R_{\mathrm{eff}}=1/\sum_\alpha w_\alpha^2$. Blue dashed lines in the Potts panels locate empirical Wolff extrema and are not equilibrium pseudo-transition assignments.}

\subsection*{\newrev{A. Representative eigen-microstates}}
\begin{figure}[p]
\centering
\includegraphics[width=0.95\linewidth]{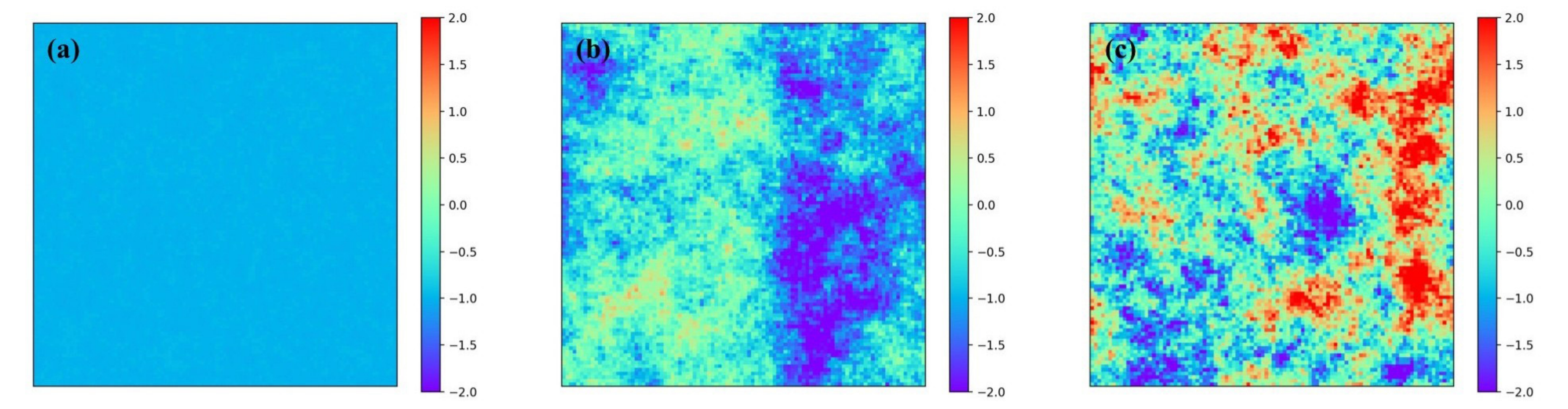}
\caption{Representative leading eigen microstates of the two-dimensional Ising model at $T=2.180$, $2.600$, and $3.000$ [(a)--(c)]. The dominant mode evolves from a nearly uniform ordered pattern to a fragmented high-temperature structure.}
\label{fig:ising_U}
\end{figure}

\newrev{Figures~\ref{fig:ising_U}, \ref{fig:RRNisingU}, and \ref{fig:potts_U} connect spectral responses with the spatial structure of representative modes. On the regular-lattice Ising model, the leading mode evolves from a nearly uniform ordered pattern to fragmented high-temperature structures. On the RRN, the leading mode captures global ordering whereas the second mode remains localized in scattered hotspots, helping explain the reduced resolvability of residual spectra. The eight-state Potts example likewise shows fragmentation of the leading mode across the main-transition window. These images clarify modal organization but are not independent transition criteria.}

\setcounter{figure}{1}
\begin{figure}[p]
\centering
\includegraphics[width=0.95\linewidth]{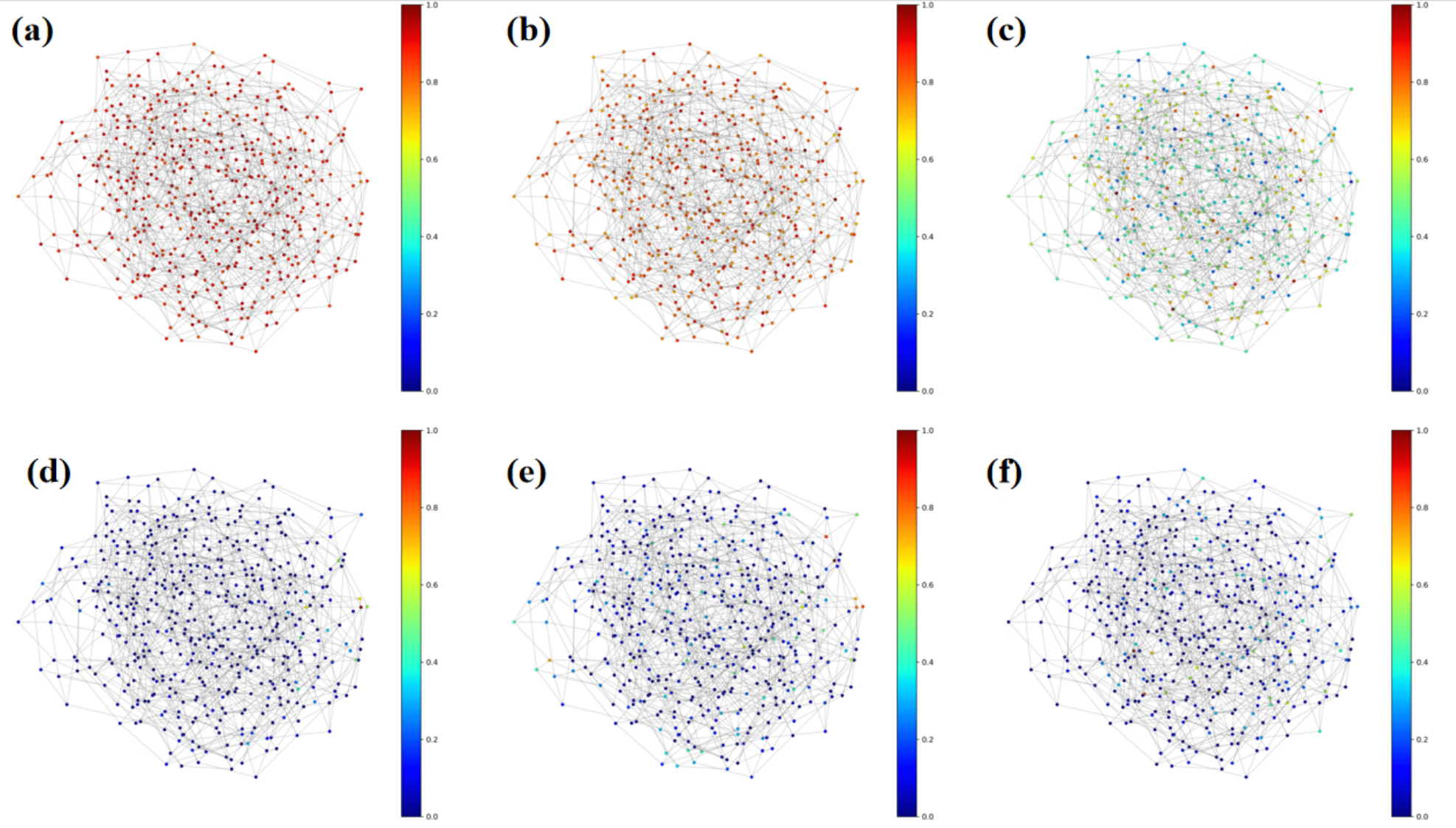}
\caption{Representative eigen microstates of the Ising model on random regular networks. Panels~(a)--(c) show the leading mode and (d)--(f) the second mode at $T=2.5$, $2.7$, and $3.25$. The second mode remains fragmented into local hotspots.}
\label{fig:RRNisingU}
\end{figure}

\setcounter{figure}{2}
\begin{figure}[p]
\centering
\includegraphics[width=0.95\linewidth]{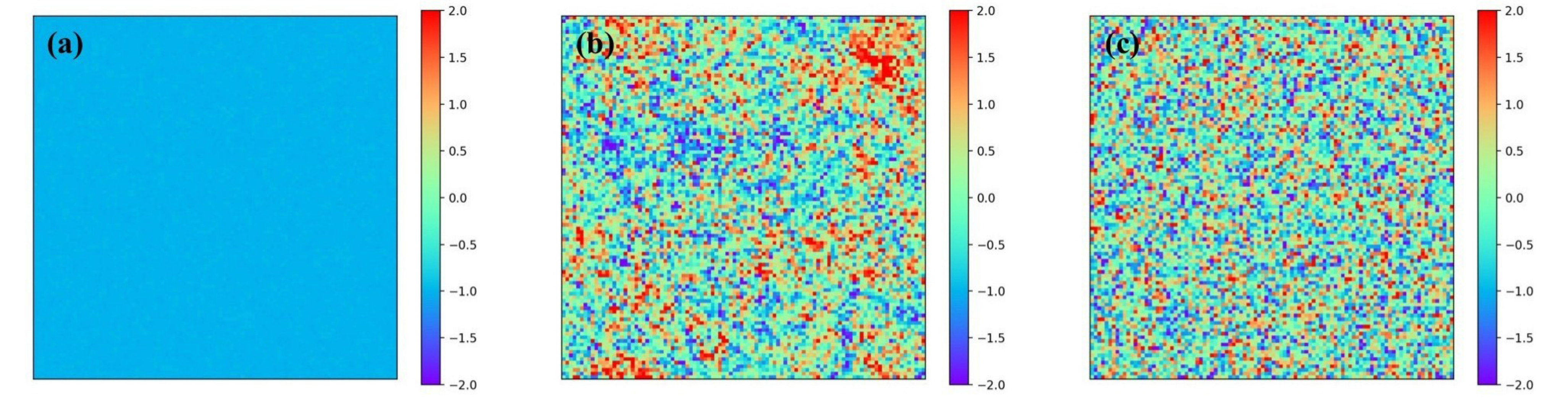}
\caption{Representative leading eigen microstates of the eight-state Potts model at $T=0.71$, $0.75$, and $0.85$ [(a)--(c)]. The dominant mode becomes fragmented across the main-transition window.}
\label{fig:potts_U}
\end{figure}
\clearpage

\subsection*{\newrev{B. Spectral moments and sampling robustness in the Ising model}}
\newrev{Figure~\ref{fig:ising_critical_moments} decomposes the critical-window response into its spectral-weight and moment components at $L=128$. The full-spectrum $R_3$ has no local minimum between $T=2.20$ and $2.40$, whereas the $k=1$ projected response has a minimum at $T=2.300$ in both Wolff and PT-Glauber spectra. The leading remaining weight and projected moments peak in the same region, showing how removing the dominant mode exposes residual-weight reorganization that need not produce a full-spectrum extremum. The comparison is limited to the tested temperatures and projection. It does not establish a mechanism for the separate high-temperature Wolff extrema, which are not correspondingly reproduced in the tested PT-Glauber full and shallow-projected spectra.}

\setcounter{figure}{3}
\begin{figure}[p]
\centering
\includegraphics[width=0.90\linewidth]{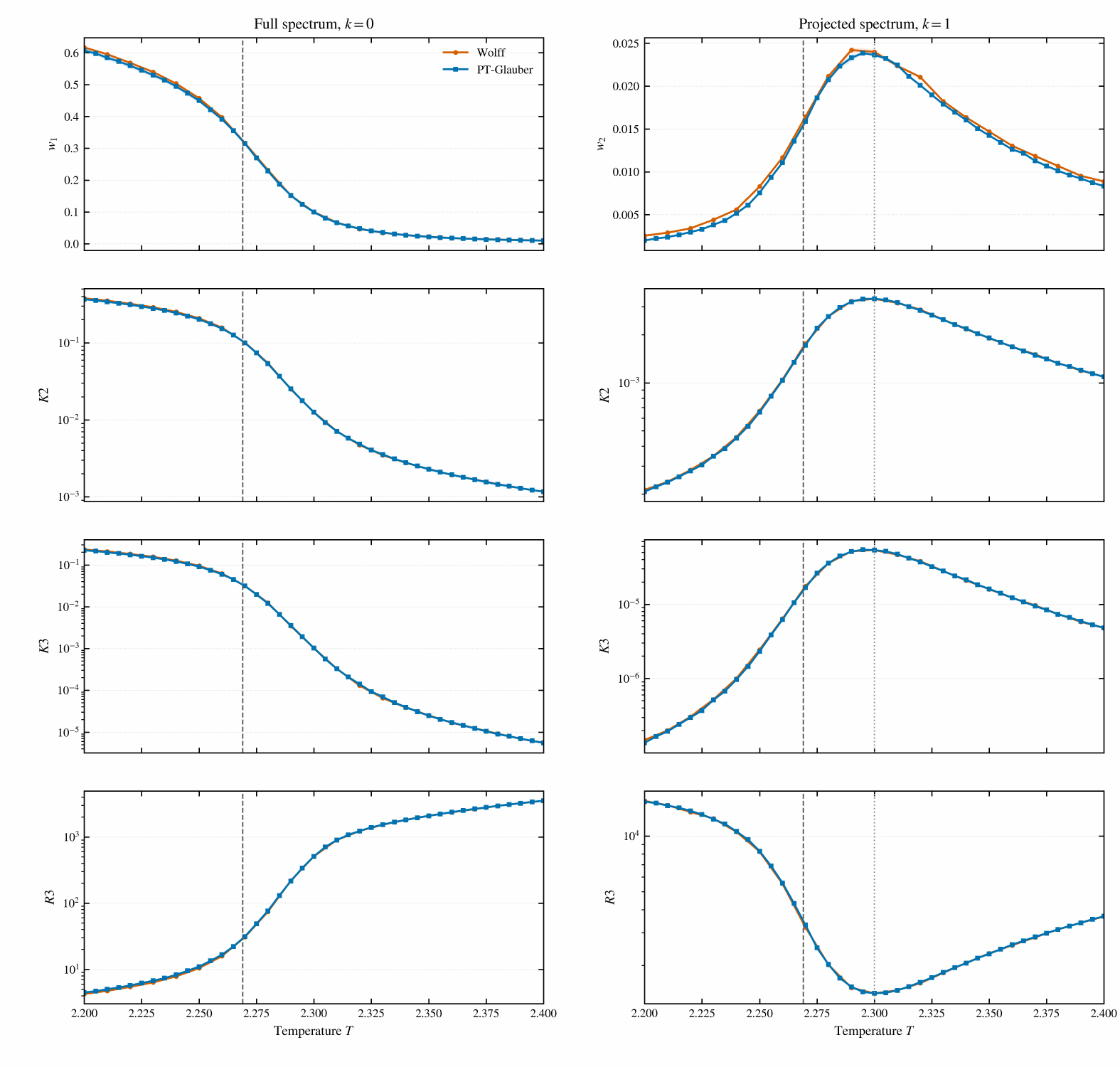}
\caption{\newrev{Critical-window spectral moments at $L=128$, averaged over five independent Wolff and PT-Glauber spectra (shaded bands: SEM). Left: full spectrum; right: $k=1$ projection. Dashed and dotted lines mark $T=2.269$ and $2.300$. The full $R_3$ has no local minimum in the displayed window, whereas $\tilde R_3^{(1)}$ has a sampling-robust minimum at $T=2.300$. The projected $\tilde w_2^{(1)}$, $\tilde K_2^{(1)}$, and $\tilde K_3^{(1)}$ identify the associated residual-weight reorganization.}}
\label{fig:ising_critical_moments}
\end{figure}
\clearpage

\subsection*{\newrev{C. Leading-mode condensation in Potts models and random regular networks}}
\newrev{Figures~\ref{fig:RRNisingW} and \ref{fig:PottsW} provide the leading-spectrum benchmarks for the RRN Ising model and the regular-lattice eight-state Potts model. The size evolution of $W_2/W_1$, the leading spectral weights, and the finite-size fits of $W_1$ establish the ordinary ordering channel. These results justify using leading-mode condensation as a reference for the higher-order analysis, but they do not by themselves resolve redistribution among subleading modes.}

\setcounter{figure}{4}
\begin{figure}[p]
\centering
\includegraphics[width=0.95\linewidth]{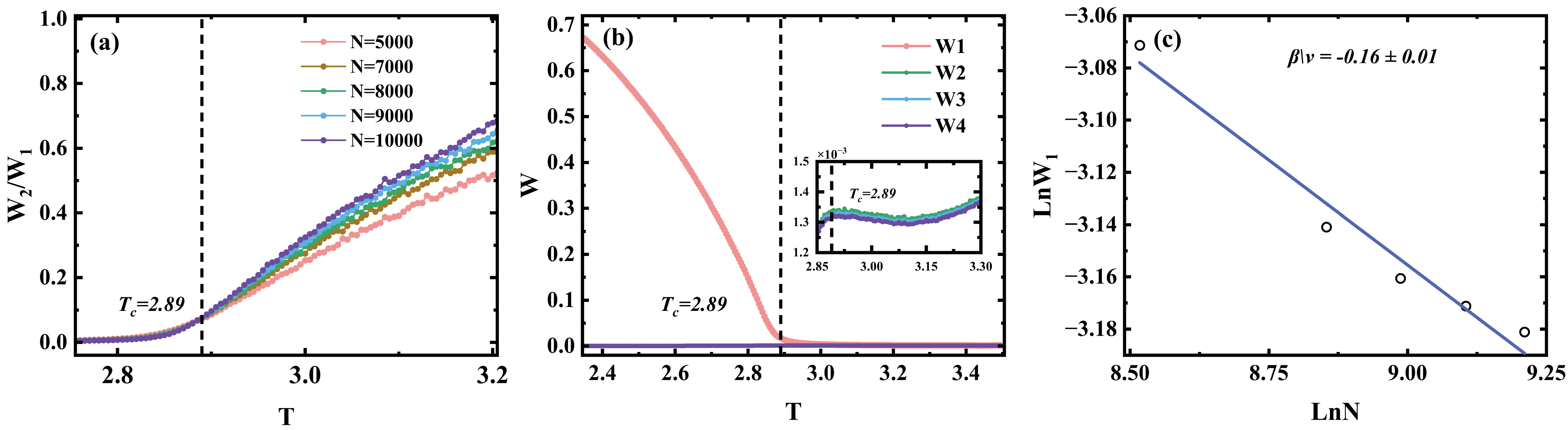}
\caption{Eigen-microstate condensation of the Ising model on random regular networks. (a)~$W_2/W_1$ versus $T$. (b)~The first four spectral weights for $N=10000$. (c)~Finite-size fit of $\ln W_1$ versus $\ln N$.}
\label{fig:RRNisingW}
\end{figure}

\setcounter{figure}{5}
\begin{figure}[p]
\centering
\includegraphics[width=0.95\linewidth]{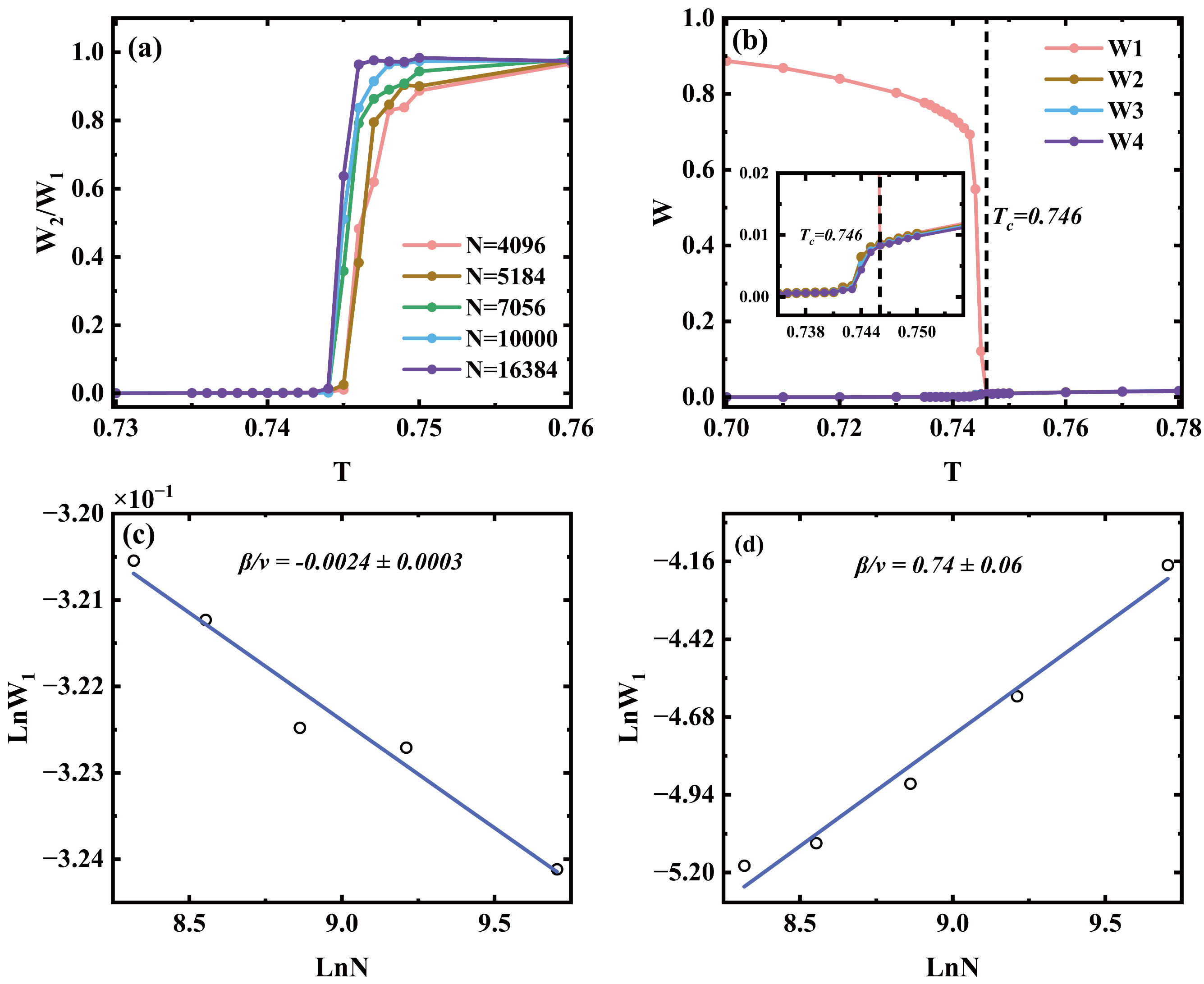}
\caption{Eigen-microstate condensation of the eight-state Potts model on a regular lattice. (a)~$W_2/W_1$ versus $T$. (b)~The first four spectral weights for $N=10000$. (c),(d)~Finite-size fits of $\ln W_1$ versus $\ln N$.}
\label{fig:PottsW}
\end{figure}
\clearpage

\subsection*{\newrev{D. Complete spectral responses of Potts models}}
\newrev{Figures~\ref{fig:Q3Potts_regular}--\ref{fig:Q8Potts_rrn} present the full-spectrum, $k=1$, and $k=100$ responses together with $R_{\mathrm{eff}}$. The regular-lattice three-state case retains an ordered-side projected feature under deeper truncation, whereas the eight-state response is compressed near the main-transition window. On RRNs, the three-state projected responses broaden and become truncation-sensitive, and the eight-state response is further compressed. These complete responses support both the applicability and the resolvability limits discussed in the main text.}

\setcounter{figure}{6}
\begin{figure}[p]
\centering
\includegraphics[width=0.95\linewidth]{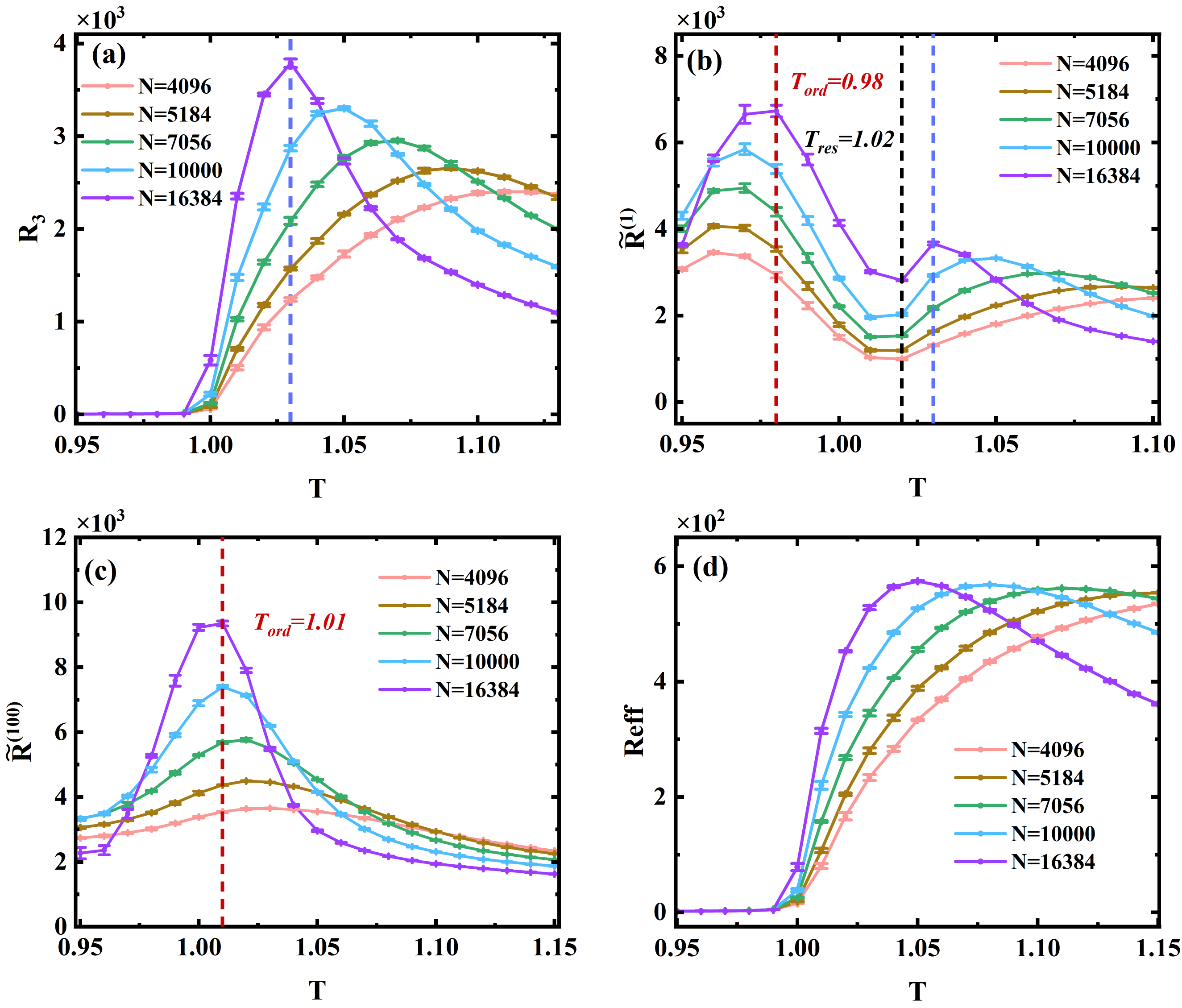}
\caption{\newrev{Complete spectral responses of the three-state Potts model on a regular lattice: (a)~full-spectrum $R_3$, (b)~$\tilde R_3^{(1)}$ after removing one mode, (c)~$\tilde R_3^{(100)}$ after removing 100 modes, and (d)~$R_{\mathrm{eff}}$. Blue dashed lines mark empirical Wolff extrema, not equilibrium transition assignments.}}
\label{fig:Q3Potts_regular}
\end{figure}

\begin{figure}[p]
\centering
\includegraphics[width=0.95\linewidth]{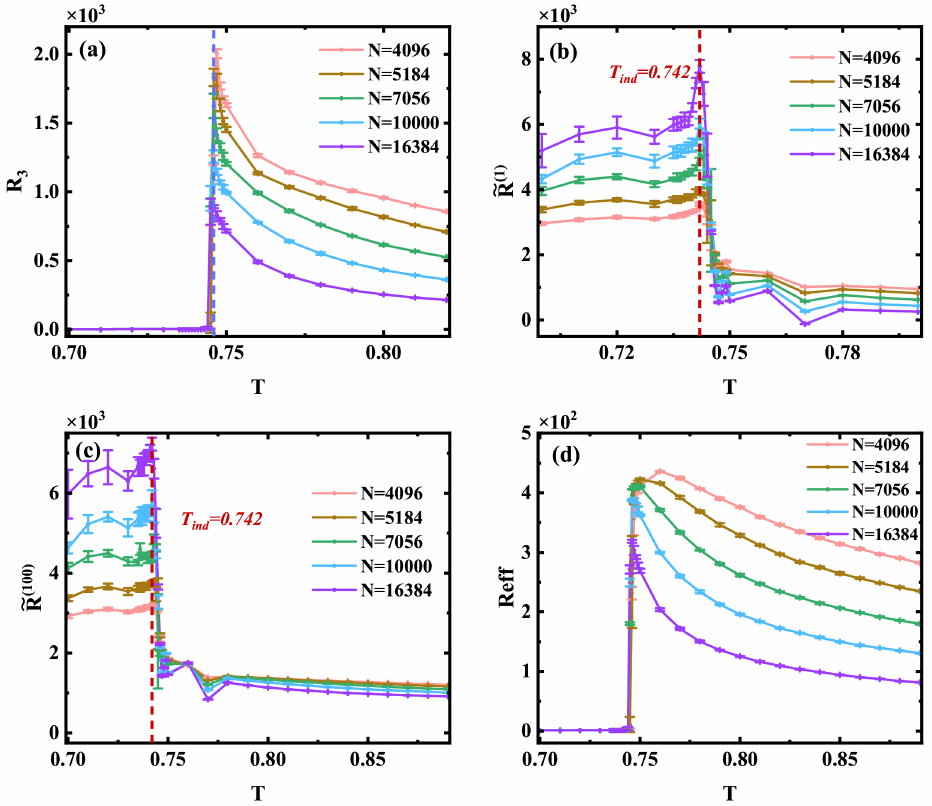}
\caption{\newrev{Complete spectral responses of the eight-state Potts model on a regular lattice: (a)~full-spectrum $R_3$, (b)~$\tilde R_3^{(1)}$, (c)~$\tilde R_3^{(100)}$, and (d)~$R_{\mathrm{eff}}$. Blue dashed lines mark empirical Wolff extrema. The responses are concentrated near the main-transition window.}}
\label{fig:Q8Potts_regular}
\end{figure}

\begin{figure}[p]
\centering
\includegraphics[width=0.95\linewidth]{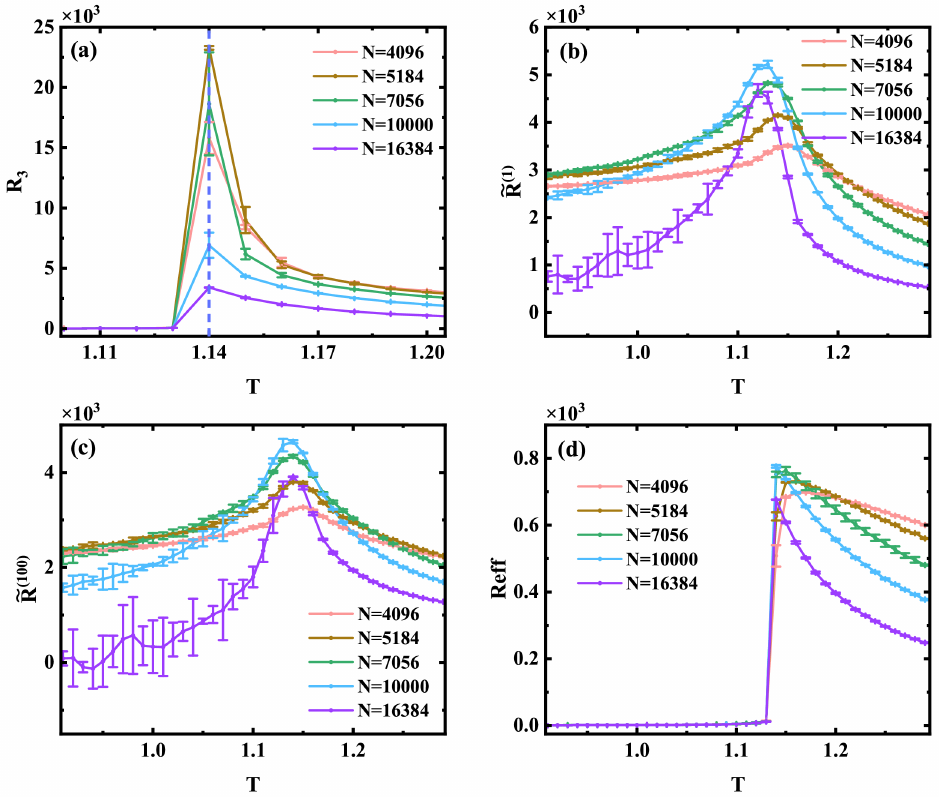}
\caption{\newrev{Complete spectral responses of the three-state Potts model on random regular networks: (a)~full-spectrum $R_3$, (b)~$\tilde R_3^{(1)}$, (c)~$\tilde R_3^{(100)}$, and (d)~$R_{\mathrm{eff}}$. Blue dashed lines mark empirical Wolff extrema. The projected responses illustrate broadening and truncation sensitivity under topological disorder.}}
\label{fig:Q3Potts_rrn}
\end{figure}

\begin{figure}[p]
\centering
\includegraphics[width=0.95\linewidth]{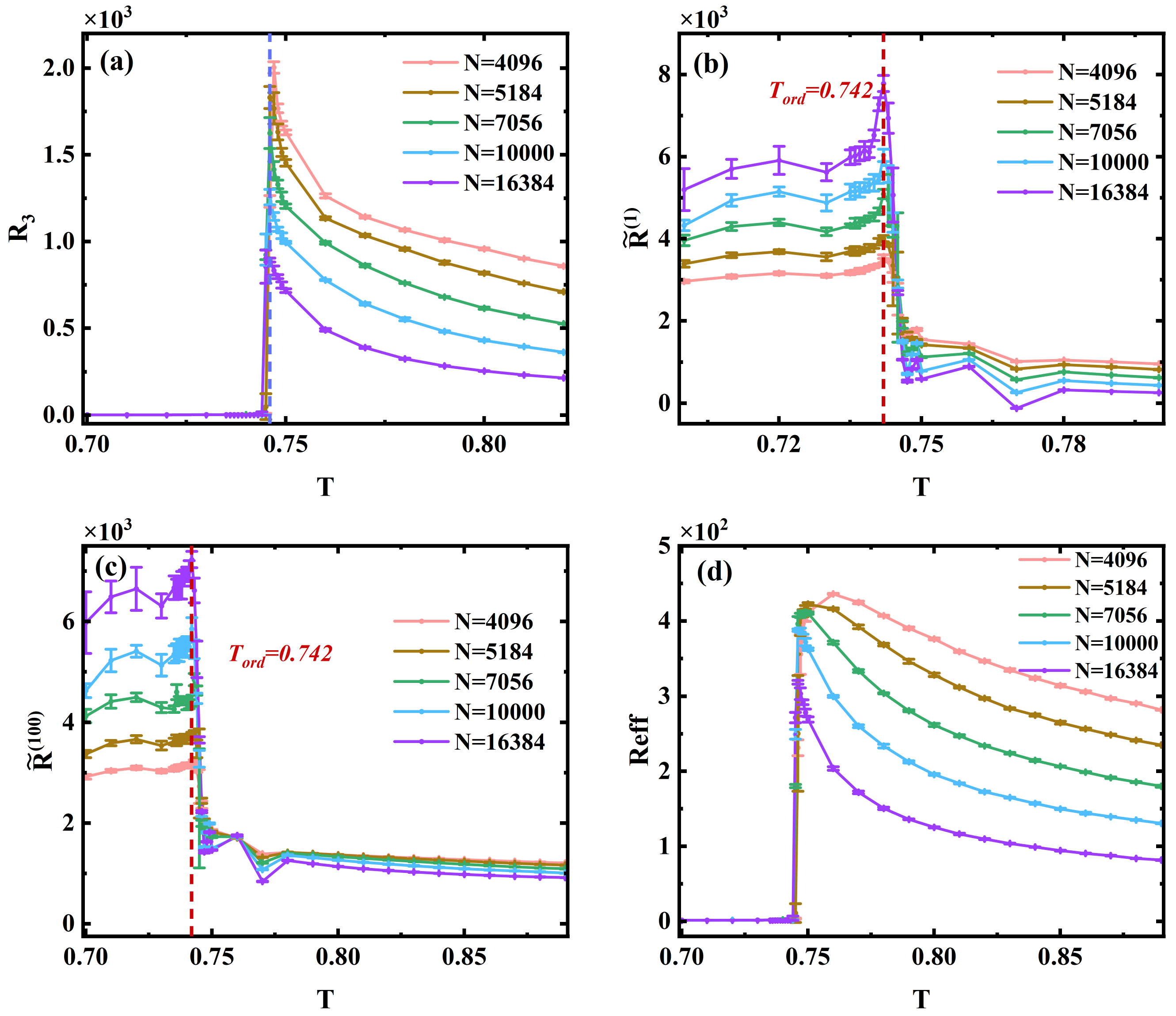}
\caption{\newrev{Complete spectral responses of the eight-state Potts model on random regular networks: (a)~full-spectrum $R_3$, (b)~$\tilde R_3^{(1)}$, (c)~$\tilde R_3^{(100)}$, and (d)~$R_{\mathrm{eff}}$. Blue dashed lines mark empirical Wolff extrema. The combined effects of larger $q$ and topological disorder concentrate the resolvable response near the main-transition window.}}
\label{fig:Q8Potts_rrn}
\end{figure}

\clearpage

\subsection*{\newrev{E. Wolff--PT-Glauber sampling cross-check}}
\newrev{Figure~\ref{fig:ising_sampling_check_supp} compares Wolff and PT-Glauber spectra for the full response and selected projections at $L=128$. It provides the sampling cross-check cited in the main text for the high-temperature extrema and the critical-window projected minimum.}

\setcounter{figure}{10}
\begin{figure}[p]
\centering
\includegraphics[width=0.85\linewidth,height=0.68\textheight,keepaspectratio]{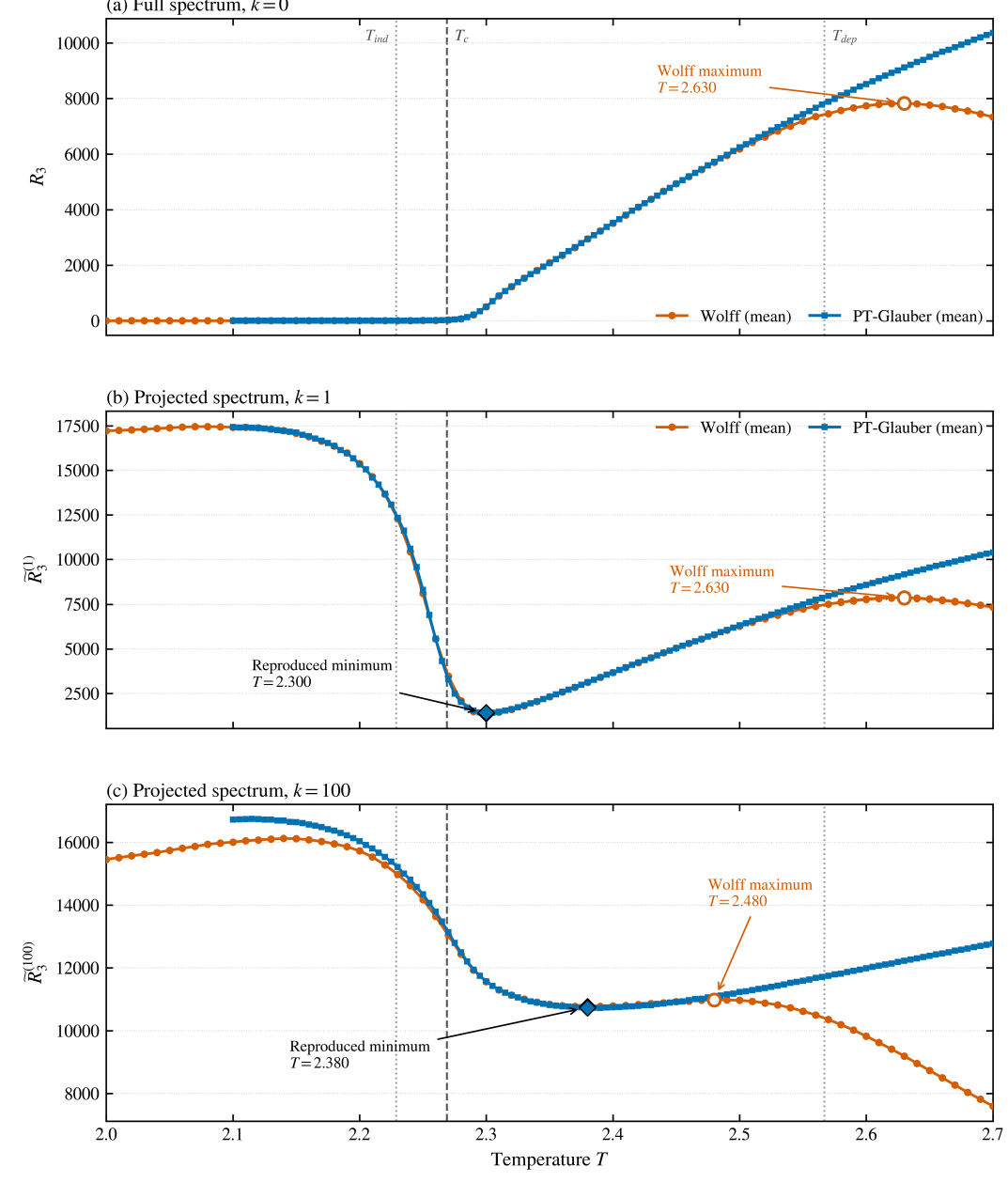}
\caption{\newrev{Wolff--PT-Glauber sampling cross-check for the full and projected responses at $L=128$. The high-temperature Wolff maxima are not reproduced as robust extrema by PT-Glauber sampling, whereas both samplers exhibit the critical-window minimum after removal of the leading mode.}}
\label{fig:ising_sampling_check_supp}
\end{figure}

\clearpage

\end{document}